\documentclass[11pt,a4paper]{article}

\usepackage{amssymb,amsmath,bm,graphicx,multirow}
\usepackage{cite,color,float}

\begin{document}
\begin{center}
\noindent{\large\textbf{Identities for Droplets with Circular Footprint on Tilted Surfaces}}
\footnote{francois.dunlop@cyu.fr,
fath@alzahra.ac.ir, 
m\_hajirahimi@azad.ac.ir,
thierry.huillet@cyu.fr
}\\~\\

Fran\c cois Dunlop$\,^{1}$, Amir H. Fatollahi$\,^2$, Maryam Hajirahimi$\,^3$, Thierry Huillet$\,^1$\\ ~\\

\textit{1) Laboratoire de Physique Th\'eorique et Mod\'elisation,\\ 
CY Cergy Paris Universit\'e, CNRS UMR 8089,\\ 95302 Cergy-Pontoise, France } 

\vskip 4mm

\textit{2) Department of Physics, Faculty of Physics and Chemistry, \\
Alzahra University, Tehran 1993891167, Iran}

\vskip 4mm

\textit{3) Physics Group, South Tehran Branch, Islamic Azad University, \\ P. O. Box 11365, Tehran 4435, Iran}

\end{center}
\vspace{.1\baselineskip}
\begin{abstract}
Exact mathematical identities are presented between the relevant parameters
of droplets displaying circular contact boundary based on flat tilted surfaces. 
 Two of the identities are derived from the force balance, and one 
from the torque balance. The tilt surfaces cover the full range of inclinations 
for sessile or pendant drops, including the intermediate case
of droplets on a wall (vertical surface).
The identities are put under test both by the available solutions of a linear response approximation at small Bond numbers
as well as the ones obtained from numerical solutions, making use of the {\sl Surface Evolver} software. The subtleties to obtain 
certain angle-averages appearing in identities by the numerical solutions 
are discussed in detail. It is argued how the identities are useful in 
two respects. First is to replace some unknown values in the Young-Laplace equation 
by their expressions obtained from the identities. Second is to use the identities to estimate the error for approximate analytical
or numerical solutions without any reference to an exact solution.
\end{abstract}

\vspace{1cm}

\noindent \textbf{Key words~: Drops, Hydrostatics, Surface tension, Young-Laplace}\\

\newpage

\section{Introduction}
Understanding the skewed shape of a sessile drop pinned on a flat
incline has a long history in Physics, starting with \cite{MO,Fr}. It
rises new asymmetrical problems compared to more studied situations
where the substrate is horizontal. In principle, such problems can be
handled while making use of the Young-Laplace nonlinear partial
differential equation, translating a balance between surface tension
forces and gravity acting on the drop. See \cite{DDH17} and references
therein, where a perturbative approach to this problem at small Bond
number was addressed when the footprint of the droplet is held fixed
and circular. More recently the case of pendant drops has also
attracted some interest, see \cite{MS88,BKM12,MWI15,MGKLSTV19} and
references therein. A perturbative approach to this problem at small
Bond number has also been addressed in \cite{cfdhs20}, in a similar
setup. An empirical relation between incline slope angle and contact
angles at the front and rear of the droplet was given by \cite{Furm},
and further studied by many authors, see \cite{DDH17}, \cite{EJ06},
\cite{DT} and references therein. It relies on an approximation of the
balance of forces equation along the substrate, at small Bond number.
Balance of forces normal to the substrate also deserves interest,
together with relations arising from the torque balance. The three-phase
contact angles (for the various azimuthal angles) of a liquid
condensed on a substrate is in direct relation with interfacial and
body forces acting on sessile or pendant drops. 

Despite the settled role of the droplets based on different surfaces, the 
cases for which there are exact solutions are rare. During more than a century, 
different numerical and analytical methods have been developed to more efficient and finer 
approach to drop's profile for cases with no exact solution. 
To evaluate or rate these numerical and analytical methods 
some criteria are needed, among which are the exact 
mathematical relations between the relevant parameters of the problem. 
An early example of these identities is the one by \cite{pitts} between 
the volume, curvature at apex, height and contact-radius of axi-symmetric 
drops on a flat horizontal surface (see also \cite{sumesh,scrfath}).
For axi-symmetric drops on curved surfaces the very same identities 
are derived in \cite{hmf}. For droplets under the combined tangential and 
normal body forces the dynamical relations between shape parameters have 
been presented in a linear approximation recently \cite{koka}.

This issue has also been studied semi-analytically in \cite{KK}, where the problem of 
understanding the  contact line evolution of  slender unpinned droplets under arbitrary 
scenarios of forces is addressed, based on experimentally observed contact lines.
Related to this point, in \cite{REKZK},  sessile droplets at different tilting angles are 
experimentally subject to varying centrifugal forces in order to explore their spreading/sliding 
behavior for different volumes and initial shapes (including non-axisymmetric). In particular, 
a test of the applicability of the Furmidge equation for the retention force is discussed.

The mathematical identities are important in another respect, that is 
reducing the initial unknown values of the problem. This in particular comes
helpful because some of these unknown values appear in the first place in 
the differential equation governing the profile of the drop. Whether one tries
a perturbative solution of the drop's profile or a numerical one, reducing the 
initially unknown values facilitates or boosts the procedure of reaching the
final result.
As an example, in \cite{scrfath} the identity is used to replace a combination 
of the unknown apex curvature and height in the Young-Laplace equation, 
in the procedure of developing a perturbative solution for lightweight drops.
As another example, in a numerical solution of axi-symmetric drops with fixed
contact angle one may use contact radius or apex height as a starting point and   
the other one as the final point. But the trouble is that both values are unknown 
in the first place, and in principle one has to search time-consumingly a two dimensional 
parameter space to find both values that match the solution.  
However, by the identity and thanks to the mentioned combination, 
one can replace the height by the radius, reducing the procedure 
to a simple one-parameter shooting method \cite{amma}. 

The main purpose of the present work is to highlight examples of 
mathematical identities for droplets with circular contact boundary 
based on flat tilted surfaces. 
The identities are derived based on the force balance along parallel and normal directions 
of the tilted surface, as well as the torque balance of the droplet. The
force balance normal to the tilted surface generalizes that of \cite{pitts,sumesh,scrfath}
for a horizontal substrate. The identity along the tilted surface is in fact an exact version of the 
approximate empirical Furmidge relation \cite{Furm}. The identity by the torque balance 
is apparently the one that is introduced here, and remains to be verified numerically.
All three identities are checked at the first order approximation of the Bond number, 
the so-called linear response ansatz \cite{DDH17}. For the identities by the force balance 
various numerical tests are provided, generated by the {\sl Surface Evolver}
as a vertex-edge-facet element software \cite{surevolv}.

The organization of the rest of the work is as follows. In Section \ref{force}
the two identities by the force-balance along and normal to the surface are
derived. In Section \ref{torque}
the identity by the torque-balance condition on the droplet is derived.
The check of all identities at linear response approximation is presented in
Section \ref{linear}. Numerical checks of force balance identities are presented
in Section \ref{numerics}. This requires 
the contact angle as a function of azimuth angle and certain averages over it, which is the subject of
Section \ref{contact}.
In Section \ref{use} the possible use of identities is illustrated.
Section \ref{conclusion} is devoted to concluding remarks.

\section{Identities by Force-Balance}\label{force}
\begin{figure}[!h]
\begin{center}
\includegraphics[width=0.6\columnwidth]{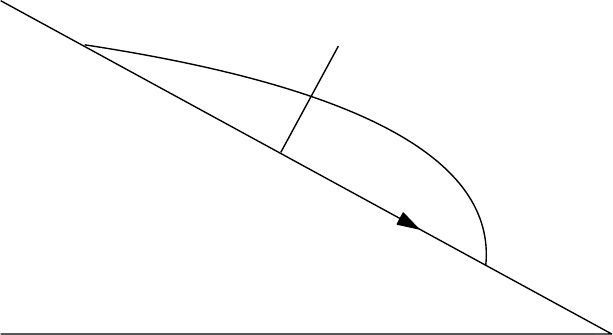}
\put(-30,3){$\alpha$}\put(-60,20){$r_0$}\put(-240,112){$-r_0$}
\put(-145,68){$c$}\put(-115,122){$z$}\put(-143,88){$h$}
\put(-127,92.5){$o$}\put(-89,34){$x$}
\caption{\small Setup of a droplet on a flat tilted surface.}
\label{fig:setup}\end{center}
\end{figure}
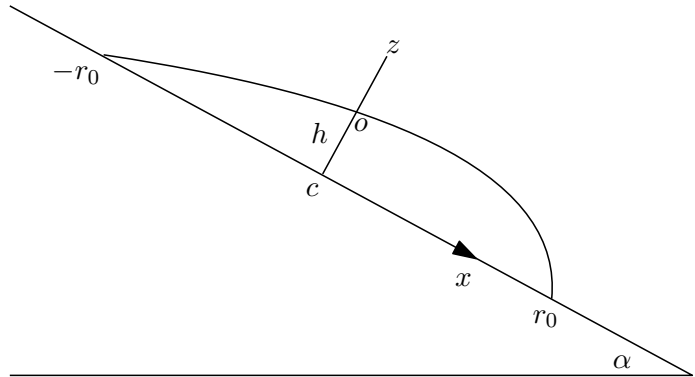

The setup for the pinned droplet on a tilted plane is as follows. The 
$z$-axis is perpendicular to the substrate and inward to the liquid, with 
$x$-axis along the slope downward. For the droplet with contact-circle of radius $r_0$,
the origin is set to be the center of circle, $c$. The angle of the substrate with 
the horizontal direction is $\alpha\in[0,\pi]$, with $\alpha=\pi$ representing the case of a drop pendant from the ceiling. The setup is summarized in Fig.~\ref{fig:setup}.
We use the polar angle $\varphi$ on the substrate, 
with $\varphi=0$ representing the $x$-axis, as usual.
The hydrostatic pressure inside the contact-circle depends only on $x$,
given by 
\begin{align}
p(x)=p_c+\rho g x \sin\alpha
\end{align}
with $\rho$ the density of the drop, and $p_c$ the pressure at the center of the contact-circle.

\subsection{Force-Balance Along Surface}
First we consider the more familiar identity, stemming from force-balance along the substrate. 
The capillary force along the $x$ direction reads:
\begin{align}
F_{x}=\gamma \int_{-\pi}^\pi r_0 \, d\varphi\, \cos\theta_\alpha(\varphi)\,\cos\varphi
\end{align}
with $\theta_\alpha(\varphi)$ the contact-angle of the liquid-substrate at 
polar angle $\varphi$ on a slope with angle $\alpha$. 
By symmetry $\theta_\alpha(\varphi)=\theta_\alpha(-\varphi)$, so we expect
the following Fourier expansion:
\begin{equation}
  \cos\theta_\alpha(\varphi)=C_0+\sum_{n=1}^\infty C_n\cos(n\varphi)
\end{equation}
where
\begin{align}
C_0&={1\over2\pi}\int_{-\pi}^\pi d\varphi\cos\theta_\alpha(\varphi)=\langle\cos\theta_\alpha(\varphi) \rangle\\
C_n&={1\over\pi}\int_{-\pi}^\pi d\varphi\cos\theta_\alpha(\varphi)\cos(n\varphi),
\quad n\ge1
\end{align}
leading to:
\begin{align}
F_{x}=\gamma \,\pi r_0\, C_1
\end{align}
The force-balance along the surface then gives:
\begin{align}\label{Fur}
mg\sin\alpha+\gamma \,\pi r_0\, C_1=0
\end{align}
with $m=\rho V$ as the mass of the droplet with volume $V$.
The dimensionless form of (\ref{Fur}) is
\begin{equation}\label{Furadim}
  2\,\mathrm{Bo}\,\sin\alpha+\pi C_1=0
\end{equation}
where $\mathrm{Bo}$ is the modified Bond (or E\"otv\"os) number taken as 
$\mathrm{Bo}=mg/(2r_0\gamma)$. Consistently with \cite{BR}, our use of 
Bo contains only the input parameters of the drop-solid system, namely volume 
of the drop, gravitational acceleration, fluid density, surface tension, and radius 
of the footprint and not data involving the dimensions of the drop unknown prior 
to the experiment (such as the height of the drop or radius of curvature at the drop apex).

The coefficient $C_1$, equals to twice the average over $\varphi$ of
$\cos\theta_\alpha(\varphi)\cos\varphi$, is remarkably linear in $\mathrm{Bo}$
as will be seen later.

In Section \ref{numerics}, in order to confront theoretical and simulation values, we will evaluate and compare to 1 the ratio
\begin{equation}\label{ratiopara}
  {\rm ratio}_\parallel=-2\,\mathrm{Bo}\,\sin\alpha/(\pi C_1)
\end{equation}

\subsection{Force-Balance Normal to Surface}
The capillary force normal to the surface (downward $z$ direction) reads:
\begin{align}
F_{z}=\gamma \int_{-\pi}^\pi r_0 \, d\varphi\, \sin\theta_\alpha(\varphi)
\end{align}
Again by symmetry we expect $\theta_\alpha(\varphi)=\theta_\alpha(-\varphi)$,
hence the Fourier expansion:
\begin{align}\label{sinth}
\sin\theta_\alpha(\varphi)=A_0+\sum_{n=1}^\infty A_n\cos(n\varphi)
\end{align}
\begin{align}
A_0&={1\over2\pi}\int_{-\pi}^\pi d\varphi\sin\theta_\alpha(\varphi)=\langle \sin\theta_\alpha(\varphi)\rangle \\
A_n&={1\over\pi}\int_{-\pi}^\pi d\varphi\sin\theta_\alpha(\varphi)\cos(n\varphi),
\quad n\ge1
\end{align}
leading to:
\begin{align}
F_{z}=2\gamma \, \pi r_0 A_0
\end{align}
The pressure force from the tilted surface is given by
\begin{align}
N&=\int_{-r_0}^{r_0} p(x) \, dA,~~~~~ dA=2\sqrt{r_0^2-x^2}dx \\
N&=\pi p_c r_0^2
\end{align}
The zero of pressure is chosen as the atmospheric pressure. Balance of forces in the normal direction leads to
\begin{align}
N=F_z+m g \cos\alpha
\end{align}
or
\begin{align}\label{normal}
\pi p_c r_0^2=\gamma \, 2\pi r_0 A_0+\rho V  g \cos\alpha
\end{align}
The pressure $p_c$ can be written in terms of the pressure at point $o$ (intersection 
point of the $z$-axis and the surface of the droplet), as follows
\begin{align}
p_c&=p_o+\rho g h \cos \alpha
\end{align}
with $h$ the $oc$ height (see Fig.~\ref{fig:setup}). This pressure,
with the use of the Young-Laplace equation, can also be written
in terms of the mean-curvature $H_0$ at point $o$
\begin{align}
p_o&=-2\gamma\, H_{0}
\end{align}
with $\gamma$ as the surface tension of liquid. 
Altogether the force-balance (\ref{normal}) reads
\begin{align}
\pi r_0^2(-2\gamma\, H_{0}\,+\rho g h \cos \alpha)
=2\gamma \pi r_0 A_0+\rho V  g \cos\alpha
\end{align}
or, after dividing by $2\pi r_0^2\gamma$,
\begin{align}\label{normal2}
-H_{0}\,+\frac{\rho g h}{2\gamma} \cos \alpha
=\frac{A_0}{r_0}+\frac{\rho V  g}{2\pi r_0^2\gamma} \cos\alpha
\end{align}
The above is a direct generalization of Eq.(12) of \cite{hmf,scrfath,amma} for a horizontal 
substrate ($\alpha=0$).

It is convenient to have, with the help of the zero-gravity copy (spherical cap) of the droplet, 
a dimensionless form of the above identity.
Using the spherical-cap radius $R_0$ and contact-angle $\theta_0$ with  
\begin{align}\label{R0}
r_0=R_0 \sin\theta_0,~~~~~~V=\frac{\pi R_0^3}{3}(1-\cos\theta_0)^2(2+\cos\theta_0)\end{align}
one defines the dimensionless quantities \cite{DDH17} 
\begin{align}
\tilde{h}=h/R_0,\qquad \tilde{H_0}=R_0 H_0 
\end{align}
The dimensionless version of (\ref{normal2}) is
\begin{equation}\label{normal3}
-\tilde H_0+{3\tilde h\sin\theta_0\,\mathrm{Bo}\cos\alpha\over\pi(1-\cos\theta_0)^2(2+\cos\theta_0)} = {A_0\over\sin\theta_0}+{\mathrm{Bo}\,\cos\alpha\over\pi\sin\theta_0}
\end{equation}

In Section \ref{numerics}, in order to confront theoretical and simulation values, we will evaluate  and compare to 1 the ratio
\begin{align}\label{ratioperp}
\textrm{ratio}_\perp=\frac{-\tilde{H}_0\,\sin\theta_0
+\displaystyle{\frac{3 \tilde{h} \sin^2\theta_0 }
{\pi(1-\cos\theta_0)^2 (2+\cos\theta_0)}} \mathrm{Bo}\,\cos \alpha}
{A_0+ \frac{1}{\pi}\,\mathrm{Bo}\,\cos\alpha}
\end{align}

\section{Identity by Torque-Balance}\label{torque}

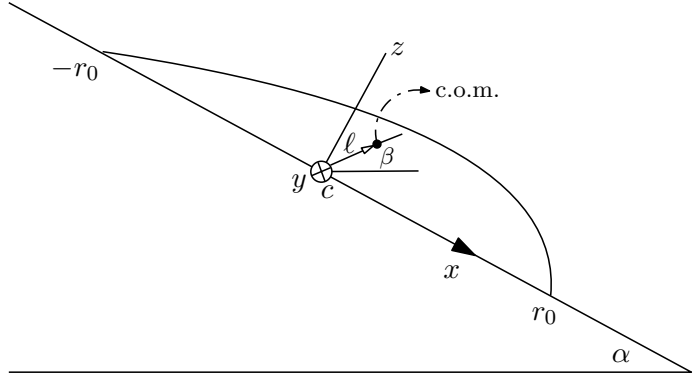
\begin{figure}[t]
\begin{center}
\includegraphics[width=0.6\columnwidth]{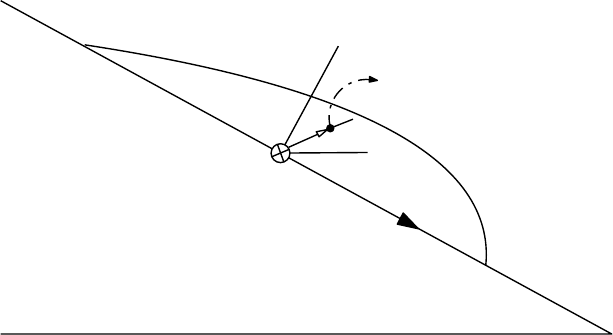}
\put(-30,4){$\alpha$}\put(-60,20){$r_0$}\put(-240,112){$-r_0$}
\put(-139,66){$c$}\put(-113,120){$z$}
\put(-93,36){$x$}\put(-150,70){$y$}
\put(-117,79){{\footnotesize $\beta$}}
\put(-130.5,83){$\ell$}
\put(-96,104){{\footnotesize c.o.m.}}
\caption{\small The geometry used for torque-balance identity.}
\label{fig:torque}
\end{center}
\end{figure}

The contribution of the weight to the torque in the $y$-direction (inward Fig.~\ref{fig:torque}) is given by 
\begin{align}
\tau_{wy}=mg\ell\cos\beta=\rho V g \ell \cos\beta
\end{align}
in which $\ell$ is the distance of the center-of-mass (c.o.m.) of the droplet
from the center $c$, and $\beta$ is the angle between the c.o.m. position vector and the horizontal direction (see Fig.~\ref{fig:torque}). 
The torque applied by the substrate upon the drop can be evaluated by integration over the element by the pressure
\begin{align}
d\tau_{py}=-x\, p(x) \, dA\,,~~~~dA=2\sqrt{r_0^2-x^2}\,dx 
\end{align}
leading to
\begin{align}
\tau_{py}&=-2\int_{-r_0}^{r_0} (p_c+\rho g x \sin\alpha) x\,\sqrt{r_0^2-x^2}\,dx\\
\tau_{py}&=-\frac{\pi}{4}\rho g r_0^4 \sin\alpha
\end{align}
The infinitesimal capillary force at polar angle $\varphi$ with contact-angle 
$\theta_\alpha(\varphi)$ reads
\begin{align}
d\vec{F}_\gamma=\gamma\, r_0\, d\varphi 
\Big(\cos \theta_\alpha(\varphi)\,\cos\varphi\, \mathbf{i}
+ \cos \theta_\alpha(\varphi)\,\sin\varphi \, \mathbf{j} 
-\sin \theta_\alpha(\varphi)\, \mathbf{k} \Big)
\end{align}
with the position vector
\begin{align}
\vec{r}=r_0 \Big(\cos\varphi\,\mathbf{i} + \sin\varphi\,\mathbf{j}\Big)
\end{align}
leading to the torque element in $y$-direction
\begin{align}
d\tau_{\gamma y}=(\vec{r}\times d\vec{F}_\gamma)_{y}=
\gamma r_0^2  \sin\theta_\alpha(\varphi)\,\cos\varphi\,d\varphi
\end{align}

The integration over the above torque element gives:
\begin{align}
\tau_{\gamma y}=\gamma r_0^2 \int_{-\pi}^\pi \sin\theta_\alpha(\varphi)\,\cos\varphi\,d\varphi
\end{align}
Again using Fourier expansion (\ref{sinth}) for $\sin\theta_\alpha(\varphi)$
one finds 
\begin{align}
\tau_{\gamma y}=\pi \gamma r_0^2\,A_1
\end{align}
All together, the balance of torques along $y$-direction gives:
\begin{align}\label{idtorque}
\rho V g\ell\cos\beta+ \pi \gamma r_0^2 A_1  = \frac{\pi}{4}\rho g r_0^4 \sin\alpha 
\end{align}
Again it is convenient to have a dimensionless form of the identity. Defining:
\begin{align}
\tilde{\ell}=\ell/R_0
\end{align}
and using (\ref{R0}) and  the relation between two Bond numbers \cite{cfdhs20}
\begin{align}\label{BBo}
\mathrm{Bo}=\pi B~\frac{(1-\cos\theta_0)^2(2+\cos\theta_0)}{6\sin\theta_0}
\end{align}
the dimensionless form of identity (\ref{idtorque}) reads:
\begin{align}\label{idtorque2}
2\,\mathrm{Bo}~ \tilde{\ell} \cos\beta + \pi \sin\theta_0 A_1= \frac{\pi}{4} B\, \sin^3\theta_0\sin\alpha
\end{align}
In Section \ref{numerics}, the following ratio is compared to 1 by the numerical 
simulations:
\begin{align}\label{ratiorot}
\mathrm{ratio}_{\bm{\circlearrowleft}}=\frac{2\,\mathrm{Bo}~ \tilde{\ell} \cos\beta + \pi \sin\theta_0 A_1}{\frac{\pi}{4} B\, \sin^3\theta_0\sin\alpha}
\end{align}

\section{Linear response}\label{linear}
\subsection{Check at Linear Response, Along Surface} 
The linear response ansatz \cite{DDH17} is in terms of the Bond number
\begin{equation}
B=\frac{\rho g R_0^2}{\gamma}
\end{equation}
where $R_0$ is the radius of the spherical cap at zero gravity, see (\ref{R0}). 
In the linear response approximation we have \cite{DDH17,cfdhs20} 
\begin{align}\label{C1}
C_1=-\frac{1}{2}\Big(\cos \theta_\alpha^\textrm{min}-\cos \theta_\alpha^\textrm{max}\Big)+O(B^2)
\end{align}
in which $\theta_\alpha^\textrm{max}=\theta_\alpha(0)$ and $\theta_\alpha^\textrm{min}=\theta_\alpha(\pi)$. 
We see that the expression (\ref{Fur})  in linear approximation is the famous Furmidge relation 
(eq.~1 of \cite{Furm}), with the constant $K=\pi/4$.

The check of (\ref{Furadim}) by explicit expressions for $\cos \theta_\alpha^\textrm{min}$ and 
$\cos \theta_\alpha^\textrm{max}$ at the linear approximation \cite{cfdhs20} is 
straightforward, once the relation between two Bond numbers (\ref{BBo}) being used.

\subsection{Check at Linear Response, Normal to Surface} \label{checknormal}
The linear response ansatz \cite{DDH17,cfdhs20} 
implies
\begin{equation}
\cos\theta_\alpha(\varphi)=\cos\theta_0+\lambda B+\mu B\cos\varphi+O(B^2)
\end{equation}
where $\theta_0$ is the uniform contact angle at $B=0$, $\lambda$ and $\mu$ are some constants 
whose values do not matter here. Then 
\begin{align}
\sin\theta_\alpha(\varphi)&=\sqrt{1-\cos^2\theta_\alpha(\varphi)} \nonumber \\
&=\sin\theta_0-\lambda B \cos\theta_0/\sin\theta_0-\mu B\cos\varphi\, \cos\theta_0/\sin\theta_0+O(B^2)
\end{align}
which implies both the average over $\varphi$
\begin{equation}
A_0=\langle\sin\theta_\alpha(\varphi)\rangle=\sin\theta_0-\lambda B \cot\theta_0+O(B^2)
\end{equation}
and the arithmetic mean between maximum and minimum
\begin{equation}
{1\over2}\Bigl(\sin\theta_\alpha^{\rm max}+\sin\theta_\alpha^{\rm min}\Bigr)=\sin\theta_0-\lambda B \cot\theta_0+O(B^2)
\end{equation}
Hence
\begin{align}\label{A0}
A_0=\frac{1}{2}\Big(\sin \theta_\alpha^\textrm{min}+\sin \theta_\alpha^\textrm{max}\Big)+O(B^2)
\end{align}

We have $\tilde{h}=1-\cos\theta_0+O(B)$ \cite{DDH17}. By the above, identity (\ref{normal3}) comes
to the form 
\begin{align}\label{last}
-\tilde{H_0}=
\frac{\sin \theta_\alpha^\textrm{min}+\sin \theta_\alpha^\textrm{max}}{2\sin\theta_0}
-\frac{B}{6}\frac{(1-\cos\theta_0)(1+2\cos\theta_0)}{1+\cos\theta_0} \cos\alpha
+O(B^2)
\end{align}

The $\tilde{H}_0$, $\sin \theta_\alpha^\textrm{min}$ and $\sin \theta_\alpha^\textrm{max}$
are read from \cite{DDH17,cfdhs20}, by which we have 
\begin{align}
-\tilde H_{0}&=1-\frac{B}{6}(1-\cos\theta_0) \cos\alpha+O(B^2)\\
\frac{1}{2}\left(\sin \theta_\alpha^\textrm{min}+\sin \theta_\alpha^\textrm{max}\right) &= \sin\theta_0-B\cos\theta_0 r'_{01}(\theta_0) \cos\alpha+O(B^2)
\label{thetaminmax}
\end{align}
with \cite{DDH17}
\begin{align}
r'_{01}(\theta_0)=-\frac{\sin\theta_0}{6} +\frac{\sin\theta_0\cos\theta_0}{3(1+\cos\theta_0)}
\end{align}
It is a simple matter to check that the quantities (\ref{A0})-(\ref{thetaminmax}) inserted into (\ref{normal3})  satisfy (\ref{normal3})  up to $O(B^2)$. 

\subsection{Check at Linear Response: Torque}
Proceeding as in Section \ref{checknormal}, the linear approximation yields 
\begin{align}
A_1&=-\frac{1}{2}\Big(\sin \theta_\alpha^\textrm{min}-\sin \theta_\alpha^\textrm{max}\Big)+O(B^2) \\
&=\frac{1}{3}B \frac{(1-\cos\theta_0)(2+\cos\theta_0)}{1+\cos\theta_0}\cos\theta_0\sin\alpha+O(B^2)
\end{align}
For a check at linear order, it is enough to insert the spherical cap droplet
values in the terms 
having $B$ or $\mathrm{Bo}$, namely the first and last terms in (\ref{idtorque2}). 
The c.o.m. of a spherical cap is known. Subtracting $R_0 \cos\theta_0$ leads to
\begin{align}
\tilde{\ell}=\frac{3(1+\cos\theta_0)^2}{4(2+\cos\theta_0)} - \cos\theta_0+O(B)
\end{align}
Also for the spherical cap the c.o.m. lays on the $z$-axis, for which we have
$\beta=\pi/2-\alpha$.
By the relation between the two Bond numbers (\ref{BBo}),
it is easy to see that identity (\ref{idtorque2}) is satisfied up to $O(B^2)$. 

\section{Numerical Check of Identities}\label{numerics}
To put test on the identities (\ref{Furadim}) and (\ref{normal3}), the solutions of
the Young-Laplace
equation are developed by the {\sl Surface Evolver} software, at different
slope angles $\alpha$'s, spherical cap contact angles $\theta_0$'s and modified Bond numbers 
$\mathrm{Bo}=mg/(2r_0\gamma)$. The identities take the form
$$
{\rm ratio}_\parallel=1,\qquad{\rm ratio}_\perp=1, \qquad 
\mathrm{ratio}_{\bm{\circlearrowleft}}=1
$$
with the ratios defined in (\ref{ratiopara}), (\ref{ratioperp}) and (\ref{ratiorot}).
In order to measure the mean curvature $\tilde H_0$ and height $\tilde h$
above the origin, we export from {\sl Surface Evolver} the list of vertices such
that in cylindrical coordinates
$\tilde r<0.06\,\tilde r_0$, that is six percent of the contact radius.
This yields between 80 and 110 vertices. 
We then fit a quadratic surface
\begin{align}
 \tilde z=a\tilde x^2+b\tilde y^2+c\tilde x+\tilde h
\end{align}
to obtain $\tilde h$ and 
\begin{align}
-\tilde H_0=(a+b(1+c^2))(1+c^2)^{-3/2}+O(B^2)
\end{align}
The measurement of $A_0$ and $C_1$ implies more work, described in
Section \ref{contact}.

\begin{figure}[H]
\begin{center}
\includegraphics[width=0.75\columnwidth]{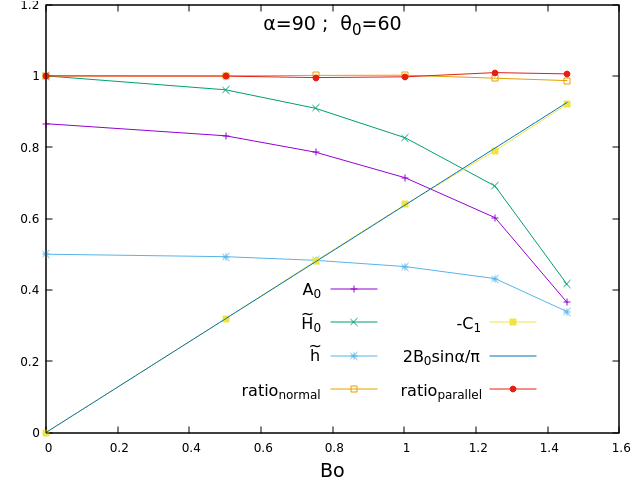}
\caption{\small Testing the identities for $\alpha=90^\circ$ and $\theta_0=60^\circ$. } \label{Ea90t60}
\end{center}
\end{figure}

\begin{figure}[H]
\begin{center}
\includegraphics[width=0.75\columnwidth]{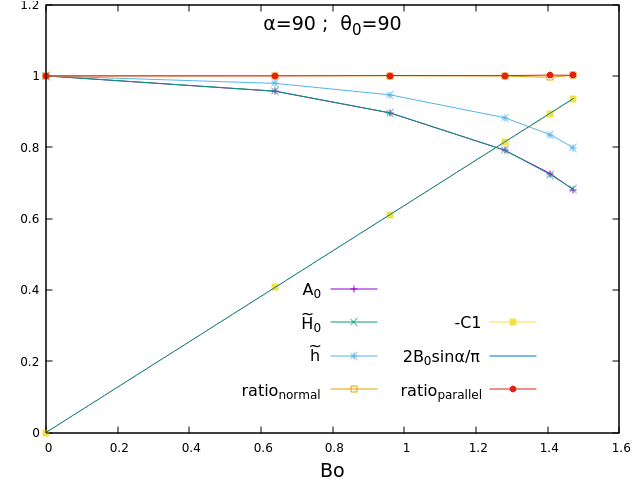}
\caption{\small Testing the identities for $\alpha=90^\circ$ and $\theta_0=90^\circ$. } \label{Ea90t90}
\end{center}
\end{figure}

\begin{figure}[H]
\begin{center}
\includegraphics[width=0.75\columnwidth]{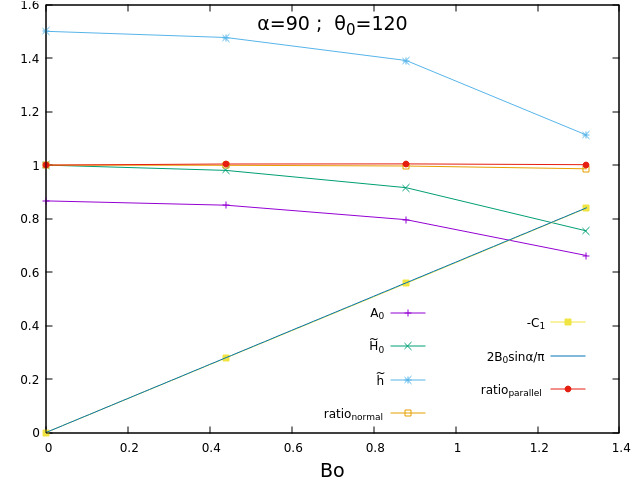}
\caption{\small Testing the identities for $\alpha=90^\circ$ and $\theta_0=120^\circ$. }
\end{center} \label{Ea90t120}
\end{figure}

In Figs. \ref{Ea90t60}-\ref{Ea90t90}-\ref{Ea90t120} we display
$A_0$, $\tilde H_0$, $\tilde h$, ratio$_\perp$, $-C_1$ and ratio$_\parallel$ as functions of $\mathrm{Bo}$
for $\alpha=90^\circ$ and $\theta_0=60^\circ,90^\circ,120^\circ$.
Tests were made with the following values 
\begin{align}
\alpha&=45^\circ, 90^\circ,135^\circ\cr
\theta_0&=60^\circ, 90^\circ, 120^\circ
\end{align}
and many different values of the $\mathrm{Bo}$ number, away from the
singularity, giving consistent values in all cases.
As a sample of the numerical values, the data for the case with 
$\alpha=90^\circ$ and $\theta_0=90$ are given in Table~1. 

\begin{table}[h!]
  \begin{center}
    \caption{Testing the identities for $\alpha=90^\circ$ and $\theta_0=90$}
    \label{tab:table1}
    \begin{tabular}{c|c|c}
Bo & ratio$_\parallel$  & ratio$_\perp$ \\
      \hline
0.63971 & 1.0000 & 1.0006\\
0.95957 & 1.0010 & 1.0001\\
1.2794 & 1.0008 & 0.99987\\
1.4073 & 1.0023 & 0.99609\\
1.4713 & 1.0017 & 1.0020
    \end{tabular}
  \end{center}
\end{table}

\begin{figure}[H]
\begin{center}
\includegraphics[width=.7\columnwidth]{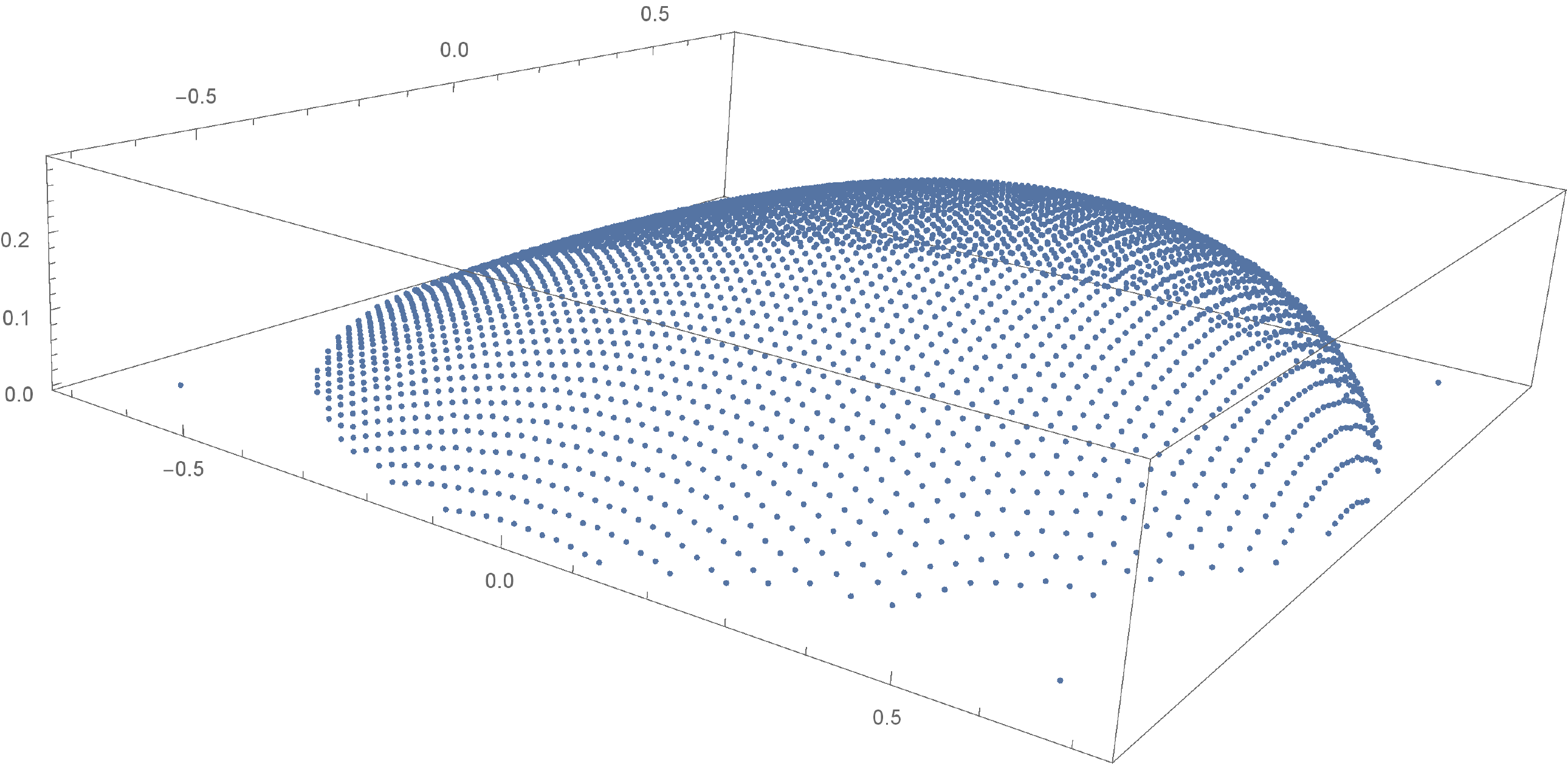}
\caption{$dx\times dy$-mesh for sample $\alpha=30^\circ$ and $\theta_0=45^\circ$,
with $\mathrm{Bo}=1.015$. Mesh size: $0.01\times 0.01$.} \label{zcm}
\end{center}
\end{figure}

\begin{figure}[H]
\begin{center}
\includegraphics[width=1.2\columnwidth]{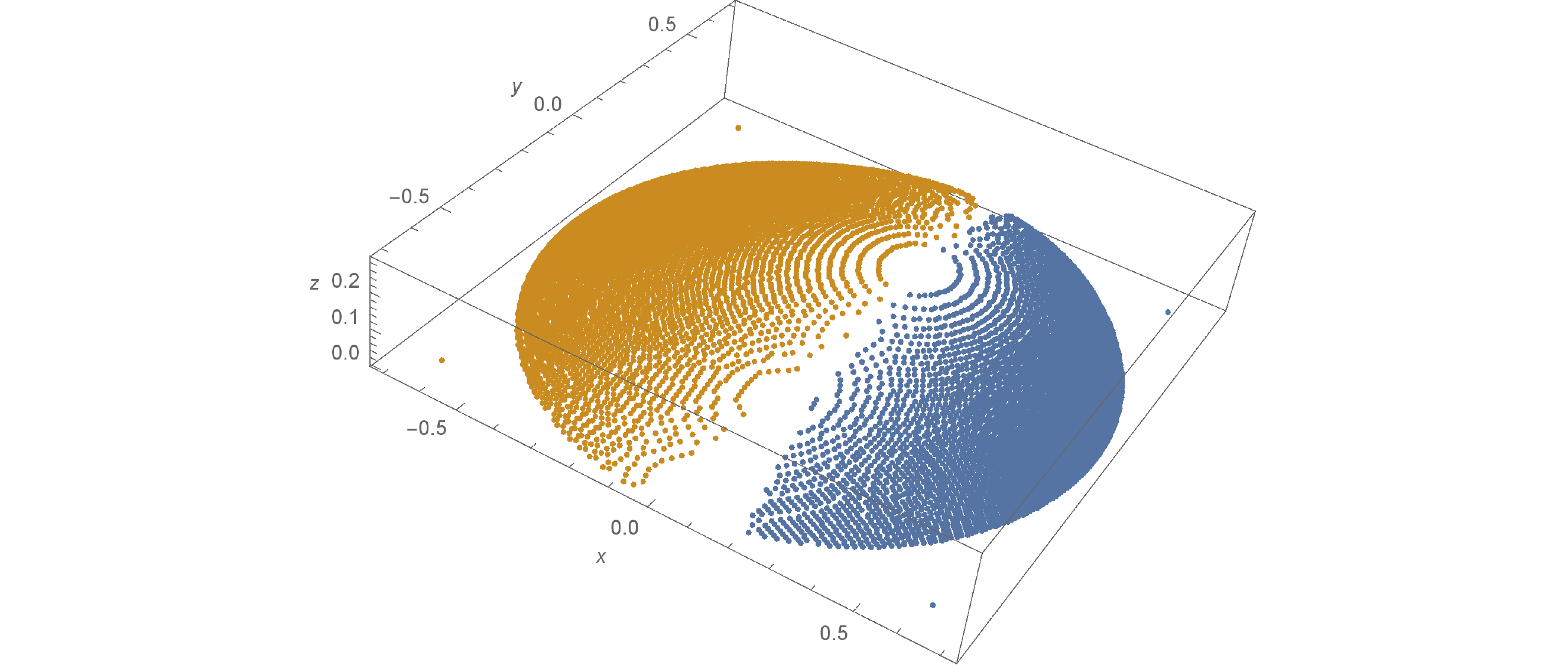}
\caption{$dy\times dz$-mesh for the sample of Fig. \ref{zcm}. Mesh size: $0.01\times 0.005$. 
Points $x>x_{zmax}$ in blue and $x<x_{zmax}$ in orange.}\label{xcm}
\end{center}
\end{figure}

In order to check the torque-balance identity numerically it is needed to 
find the c.o.m. of the droplet, for which out of the irregular distribution of 
vertices one has to extract regular $dx\times dy$-mesh
and $dy\times dz$-mesh, for calculating $\tilde{z}_{cm}$ and $\tilde{x}_{cm}$, respectively.
The result of making meshes for one of the samples
is given in Figs. \ref{zcm} and \ref{xcm}.
The values of $\tilde{l}$ and angle $\beta$ are then obtained by
\begin{align}
\beta&=\tan^{-1} \frac{z_{cm}}{x_{cm}}-\alpha \\
\tilde{\ell} &= \sqrt{\tilde{x}_{cm}^2+\tilde{z}_{cm}^2}
\end{align}
To obtain the average-value $A_1$ the azimuth-angle is needed; the computation 
is postponed to Section \ref{contact}.
The results of torque identity checks are summarized in Table \ref{torqtab}.
\begin{table}[h!]
  \begin{center}
    \caption{Testing the torque balance identity}
    \label{torqtab}
 \begin{tabular}{c|c|c|c}
$\alpha$ & $\theta_0$ &Bo & $\mathrm{ratio}_{\bm{\circlearrowleft}}$ \\
      \hline
$30^\circ$ & $45^\circ$ & 1.02 & 0.999\\
$30^\circ$ & $90^\circ$ & 0.512 & 1.01 \\
$45^\circ$ & $30^\circ$ & 0.453 &  0.999 \\
$60^\circ$ & $60^\circ$ & 0.401 &  1.00
    \end{tabular}
  \end{center}
\end{table}

\section{Contact Angle as Function of Azimuth}\label{contact}
In order to check the {\sl Surface Evolver} simulations against the exact
identities, we have to compute $C_0$ and $C_1$, or the average over
$\varphi\in(0,2\pi)$ of $\sin\theta(\varphi)$ and the average over
$\varphi\in(0,2\pi)$ of $\cos\varphi\cos\theta(\varphi)$, where
$\theta(\varphi)$ is the contact angle at azimuth $\varphi$.
A fluid interface in contact with a solid surface has a contact angle
$\theta\in(0,\pi)$ which may vary along the contact line. The solid surface
must be smooth at the macroscopic scale, so that a unique normal vector
is defined at every point.

We have performed the {\sl Surface Evolver} simulations with Dirichlet boundary
conditions: the displacement is zero on the contact line, a fixed circle of
radius $r_0$. Physically, other than a droplet on or below an incline, it may
represent a pocket of liquid made with an elastic
membrane fixed onto a circular metallic wire, opening to a reservoir of
liquid. Initially, at zero gravity, the pocket has the shape of a spherical cap
of contact angle $\theta_0$, giving the desired volume. The pocket lies entirely
on one side, say $\{z\ge0\}$, of the plane containing the contact line circle.
Upon switching on the gravity, the pocket deforms as a
solution of the Young-Laplace equation under constant volume constraint and the
given Dirichlet boundary conditions. Eventually the interface or membrane may
go partly into $\{z<0\}$ region.

\begin{figure}[H]
\begin{center}
\includegraphics[width=0.75\columnwidth]{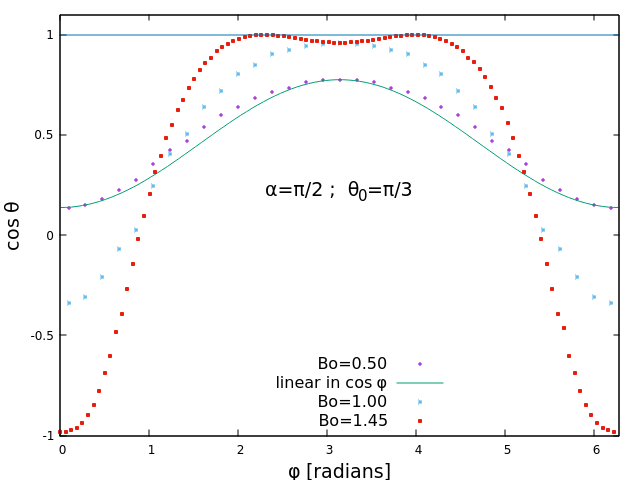}
\caption{\small Cosine of contact angle $\theta$ as function of azimuth $\varphi$
for $\alpha=90^\circ$ and $\theta_0=60^\circ$ at Bond numbers
  $\mathrm{Bo}=0.50,\,1.00,\,1.45$. Linear response approximation is plotted at $\mathrm{Bo}=0.50$.}
\label{cosa90t60}\end{center}
\end{figure}

\begin{figure}[H]
\begin{center}
\includegraphics[width=0.75\columnwidth]{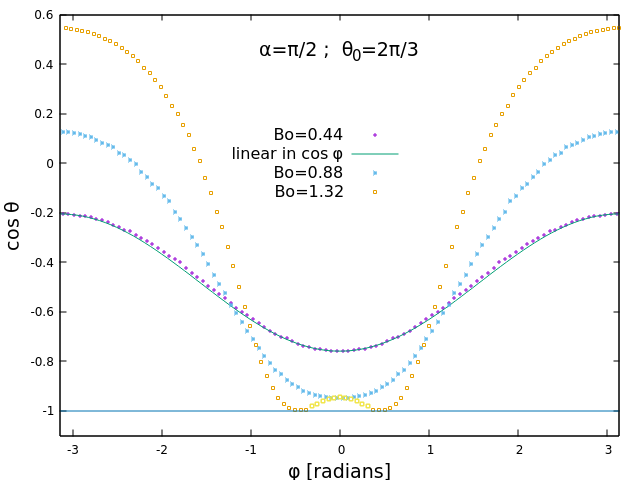}
\caption{\small Cosine of contact angle $\theta$ as function of azimuth $\varphi$
for $\alpha=90^\circ$ and $\theta_0=120^\circ$ at Bond numbers
  $\mathrm{Bo}=0.44,\,0.88,\,1.32$. Linear response approximation is plotted at $\mathrm{Bo}=0.44$.}
\label{cosa90t120}\end{center}
\end{figure}

Of course, so far as the interface remains in $\{z\ge0\}$, it represents as well
a liquid drop on or below a solid plane substrate, with contact line pinned on
the circle and contact angle $\theta\in(0,\pi)$ varying along the circle.
However, where the interface goes into $\{z<0\}$, representing a pocket, a
contact angle $\theta>\pi$ or $\theta<0$ will be found. This occurs e.g. for
$\alpha=\pi/2,\,\theta_0=\pi/3,\,\mathrm{Bo}=1.46$ around $\varphi=\pi$
(Fig. \ref{cosa90t60})
and for $\alpha=\pi/2,\,\theta_0=2\pi/3,\,\mathrm{Bo}=1.32$ around $\varphi=0$
(Fig. \ref{cosa90t120}).
At such parameters, a droplet on the incline would have unpinned, dewetting
from the top around $\varphi=\pi$ or overflowing at the bottom  around
$\varphi=0$. The Young-Laplace equation then has to be solved with moving
boundary, which we have not done.

The figures also show the linear response approximation, namely $\cos\theta$
as a linear function of $\cos\varphi$, at the smallest Bond
number for which data are displayed.

As another noticeable feature, it is observed that to some approximation, for each choice of
$\alpha$ and $\theta_0$, there is an azimuth $\varphi$ at which
$\theta\simeq\theta_0\ \forall ~\mathrm{Bo}$.

The number of mesh vertices in the {\sl Surface Evolver} simulation is up to
700000, with finer mesh near the contact line and finer mesh also
in the more delicate cases
where the Bond number Bo approaches the instability threshold.
In order to measure the contact angle $\theta$ as function of the azimuth
$\varphi$, we export from {\sl Surface Evolver} the list of vertices such that
$|\tilde z|<0.03\,(1-\cos\theta_0)$, that is three percent of the height of the
drop at zero Bond number. This yields between 1300 and 20000 vertices. We divide
the range of azimuth into a hundred equal sectors. Each sector contains between
four and six hundred vertices.
In each sector we fit with {\sl gnuplot} a surface containing the 
contact line in the form, in cylindrical cordinates,
$$
r(z)=r_0-z\cot\theta+az^2\qquad\hbox{or}\qquad z(r)=-(r-r_0)\tan\theta+b(r-r_0)^2
$$
The fit is performed, independently in each sector, in terms of $\cot\theta$
and $a$ or $\tan\theta$ and $b$. 

\section{Uses of Identities}\label{use}
\subsection{In First Place: An Illustrative Example}
The identity by force-balance in normal direction can be used in the first 
place in the Young-Laplace equation. It helps to replace some unknown values in 
the equation from the beginning. This has been used in \cite{scrfath} to 
develop a perturbation solution and in \cite{amma} to reduce the 
numerical procedure to a simple shooting method,
for sessile drops on horizontal surface with fixed contact-angle. 
Here we show that the identity can be used to replace the unknown 
quantities in the Young-Laplace equation for drops with 
fixed contact-radius (pinned drops) on tilted surfaces as well. 

First one recognizes that the Young-Laplace equation can be written in the form:
\begin{align}
- 2\gamma H_{0}+\rho g (h-z)\cos\alpha=- 2\gamma H
\end{align}
in which $H_{0}$ and $h$ are unknown in the first place. 
Now by the identity (\ref{normal2}) one has for the combination as follows
\begin{align}
-2\gamma H_{0}\,+\rho g h \cos \alpha
=2\gamma \frac{A_0}{r_0}+\frac{\rho V  g}{\pi r_0^2} \cos\alpha
\end{align}
bringing the Young-Laplace to the form:
\begin{align}
2\gamma \frac{A_0}{r_0}+\frac{\rho V  g}{\pi r_0^2} \cos\alpha
-\rho g z\cos\alpha=- 2\gamma H
\end{align}
Now in the LHS only $A_0$ is unknown. To develop a perturbative 
analytical solution, such as in \cite{DDH17}, the above form has the advantage that
it is involved by less to be calculated values. At linear order of $B$ number, 
the above form finds its full advantage, because at the beginning even $A_0$ is known by 
the linear ansatz (\ref{A0}).
As mentioned earlier, the numerical solution in reduced form helps lowering the parameters space dimension \cite{amma}.

\subsection{Estimation of Error by Identities} 
\begin{figure}[H]
\begin{center}
\includegraphics[width=0.6\columnwidth]{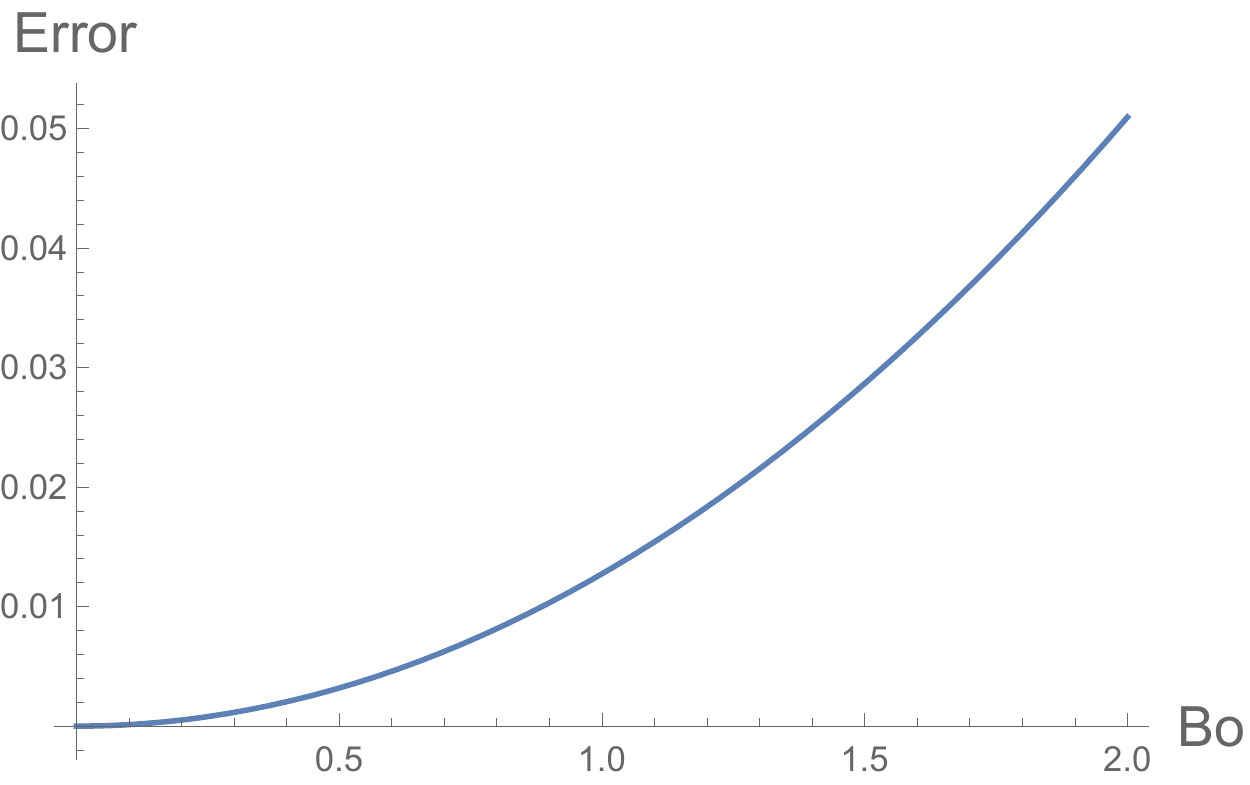}
\caption{\small The estimation of error of $-\tilde{H}_0$ vs. $\mathrm{Bo}$ for $\alpha=45^\circ$ and $\theta_0=120^\circ$.}
\label{ErrBo}
\end{center}
\end{figure}

As mentioned earlier, the exact mathematical relation may be used to make
an estimation error for an analytical or numerical approach. As
an example, we use identity (\ref{normal3}) by force-balance normal to the substrate 
to estimate the error of $-\tilde{H}_0$ by the linear approximation of \cite{cfdhs20}, 
that is expected to be of second order in $B$. 
First let us rearrange the identity in the form that the apex-curvature 
would be calculated by other values, namely:
\begin{align}
\left[-\tilde{H}_0\right]_\mathrm{Identity}=-
{3\tilde h\sin\theta_0\,\mathrm{Bo}\cos\alpha\over\pi(1-\cos\theta_0)^2(2+\cos\theta_0)}
+ {A_0\over\sin\theta_0}+{\mathrm{Bo}\,\cos\alpha\over\pi\sin\theta_0}
\end{align}
The above gives the value of $-\tilde{H}_0$ required to hold the force-balance condition.
It is seen earlier in Section \ref{checknormal} that, when only the first order is kept the above equality holds. Now
\begin{align}
\tilde{h}=1-\cos\theta_0+\frac{B}{6}\left(1-\cos\theta_0+2\ln\frac{1+\cos\theta_0}{2}\right)\cos\alpha+O(B^2)
\end{align}
When expressed based on the number $\mathrm{Bo}$ via relation (\ref{BBo}) (also 
(8) of \cite{cfdhs20}), the estimation of error is given by:
\begin{align}
\frac{\left[-\tilde{H}_0\right]_\mathrm{identity}-\left[-\tilde{H}_0\right]_\mathrm{linear~sol}}{\tilde{R}_0^{-1}}=&
\frac{-\mathrm{Bo}^2}{3\pi^2}\frac{\sin^2\theta_0}{(1-\cos\theta_0)^4(2+\cos\theta_0)^2}\times
\cr 
&\left(1-\cos\theta_0+2\ln\frac{1+\cos\theta_0}{2}\right) \cos^2\alpha
\end{align}
in which we have used $\tilde{R}_0=1$.
The remarkable observation by \cite{cfdhs20} is that, for relatively large values of the Bond number
$\mathrm{Bo}$ the linear approximation solution is quite close to the numerical solution. 
In Fig. \ref{ErrBo},  we see that for Bond numbers up to $\mathrm{Bo}=2.0$ the error is less than $5\%$.

\section{Conclusion}\label{conclusion}
The main concern of the present work is to highlight  
exact mathematical identities for droplets, even in cases where an
exact Young-Laplace profile solution is not available. 
As a special case,  droplets with circular contact boundary based on flat tilted surfaces
are considered. 
Two of the identities are derived as the requirements of the force-balance, and one 
of the torque-balance. 
The identities involve some relevant values of droplets. On the side of output values, some are 
available once an analytical or numerical solution would be given,
for instance the curvature at apex or the vertical-height. Some of these output values
appearing in the identities are somehow indirectly available, such as certain 
azimuthal-angle averages, or the location of the center-of-mass of the droplet.
The identities are put under test both by the available solutions of a linear response approximation
as well as the ones obtained from exact numerical solutions.  
We stress that the problem analyzed  here should not only be seen as an abstract theoretical construction. It is a real problem having an experimental counterpart e.g. putting the droplet on a disc shaped asperity on the substrate.

\vspace{5mm}
\textbf{Acknowledgement}: 
The work by A.H. Fatollahi is supported by the Research Council of Alzahra University.
\bibliographystyle{RS}
\bibliography{identities}

\begin{thebibliography}{99}

\bibitem{MO}
Macdougall G, Ockrent C. 1942  Surface energy relations in liquid/solid systems
  I. The adhesion of liquids to solids and a new method of determining the
  surface tension of liquids. {\em Proc. R. Soc. London Ser. A}
  \textbf{180(981)}, 151.

\bibitem{Fr}
Frenkel YI. 1948  On the behavior of liquid drops on a solid surface 1. The
  sliding of drops on an inclined surface. {\em Zh. Eksp. Teor. Fiz.}
  \textbf{18}, 659 (Translated by V. Berejnov:
  https://arxiv.org/abs/physics/0503051).

\bibitem{DDH17}
{De Coninck} J, Dunlop F, Huillet T. 2017  Contact angles of a drop pinned on
  an incline. {\em Phys. Rev. E} \textbf{95}, 052805.

\bibitem{MS88}
Milinazzo F, Shinbrot M. 1988  A numerical study of a drop on a vertical wall.
  {\em J. Colloid. Interface Sci.} \textbf{121}, 254.

\bibitem{BKM12}
Bhutani G, Khandekar S, Muralidhar KV. 2012  Contact angles of pendant drops on
  rough surfaces. {\em Proceedings of the Thirty Ninth National Conference on
  Fluid Mechanics and Fluid Power, Gujarat, India}.

\bibitem{MWI15}
de~la Madrid R, Whitehead, T., Irwin G. 2015  Comparison of the Lateral
  Retention Forces on Sessile and Pendant Water Drops on a Solid Surface. {\em
  Am. J. Phys.} \textbf{83}, 531--538.

\bibitem{MGKLSTV19}
de~la Madrid R, Garza F, Kirk J, Luong H, Snowden L, Taylor J, Vizena B. 2019
  Comparison of the Lateral Retention Forces on Sessile, Pendant, and Inverted
  Sessile Drops. {\em Langmuir} \textbf{35(7)}, 2871--2877.

\bibitem{cfdhs20}
{De Coninck} J, Fernandez-Toledano J, Dunlop F, Huillet T, Sodji A. 2021  Shape
  of pendent droplets under a tilted surface. {\em Physica D: Nonlinear
  Phenomena} \textbf{415}, 132765.

\bibitem{Furm}
Furmidge C. 1962  Studies at phase interfaces I. The sliding of liquid drops on
  solid surfaces and a theory for spray retention. {\em J. Colloid Science}
  \textbf{17}, 309.

\bibitem{EJ06}
ElSherbini A, Jacobi A. 2006  Retention forces and contact angles for critical
  liquid drops on non-horizontal surfaces. {\em J. Colloid Interface Sci.}
  \textbf{299}, 841.

\bibitem{DT}
{Dussan V.} EB, Chow RTP. 1983  On the ability of drops or bubbles to stick to
  non-horizontal surfaces of solids. {\em J. Fluid Mech.} \textbf{137}, 1--29.

\bibitem{pitts}
Pitts E. 1974  The stability of pendent liquid drops. Part 2. Axial symmetry.
  {\em J . Fluid Mech.} \textbf{63}, 487--508.

\bibitem{sumesh}
Sumesh PT, Govindarajana R. 2010  The possible equilibrium shapes of static
  pendant drops. {\em The Journal of Chemical Physics} \textbf{133}, 144707.

\bibitem{scrfath}
Fatollahi AH. 2012  On the shape of a lightweight drop on a horizontal plane.
  {\em Physica Scripta} \textbf{85(4)}, 045401.

\bibitem{hmf}
Hajirahimi M, Mokhtari F, Fatollahi AH. 2015  Exact identities for sessile
  drops. {\em Appl. Math. Mech.} \textbf{36(3)}, 293.

\bibitem{koka}
Kostoglou M, Karapantsios T. 2020  An Analytical Two-Dimensional Linearized
  Droplet Shape Model for Combined Tangential and Normal Body Forces. {\em
  Colloids Interfaces} \textbf{4}, 35.

\bibitem{KK}
Kostoglou M, Karapantsios T. 2019  Contact Angle Profiles for Droplets on
  Omniphilic Surfaces in the Presence of Tangential Forces. {\em Colloids
  Interfaces} \textbf{3}, 60.

\bibitem{REKZK}
Ríos-López I, Evgenidis S, Kostoglou M, Zabulis X, Karapantsios T. 2018
  Effect of initial droplet shape on the tangential force required for
  spreading and sliding along a solid surface. {\em Colloids and Surfaces A:
  Physicochemical and Engineering Aspects} \textbf{549}, 164--173.

\bibitem{amma}
Fatollahi AH, Hajirahimi M. 2013  Making sessile drops easier. {\em
  arXiv:1304.6366}.

\bibitem{surevolv}
Brakke K. 2013  Surface Evolver Manual. {\em Susquehanna University, \\
  http://facstaff.susqu.edu/brakke/evolver/evolver.html}.

\bibitem{BR}
Berim G, Ruckenstein E. 2020  Bond Number Revisited: Axisymmetric Macroscopic
  Pendant Drop. {\em Langmuir} \textbf{36}, 6512--6520.

\end{thebibliography}
\end{document}